\documentclass[12pt,draftcls,onecolumn]{IEEEtran}
\usepackage{graphicx}
\usepackage{amsmath}
\usepackage{ntheorem}
\usepackage[noend]{algpseudocode}
\usepackage{algorithmicx,algorithm}
\usepackage{amssymb}
\usepackage{epstopdf}
\usepackage{setspace}
\usepackage{color}
\usepackage{array}
\usepackage{bm}
\usepackage{url}
\usepackage{amsfonts}
\usepackage{stfloats}
\usepackage[numbers,sort&compress]{natbib}
\usepackage{bm}
\usepackage[caption=false,font=normalsize,labelfont=sf,textfont=sf]{subfig}
\usepackage{textcomp}
\usepackage{verbatim}

\begin{document}

\title{Joint Resource Allocation and Trajectory Design for Resilient Multi-UAV Communication Networks}

\author{
Linghui~Ge,~\IEEEmembership{Student Member, IEEE,}
Xiao~Liang,~\IEEEmembership{Member, IEEE,}
Hua~Zhang,~\IEEEmembership{Member, IEEE,}
Peihao~Dong,~\IEEEmembership{Member, IEEE,}
Jianxin~Liao,~\IEEEmembership{Member, IEEE,}
Jingyu~Wang,~\IEEEmembership{Member, IEEE}
}

\maketitle

\begin{abstract}
In contrast to terrestrial wireless networks, dynamic Unmanned Aerial Vehicle (UAV) networks are susceptible to unexpected link failures arising from UAV breakdowns or the depletion of its batteries. Drastic user rate fluctuations and sum rate drops can occur due to the unexpected UAV link failures. Previous research has focused primarily on re-establishing these links to maintain service continuity, while neglecting overall system performance, including sum rate and user rate fluctuations. This letter proposes a resilient UAV network design utilizing the modern portfolio theory (MPT), which jointly optimizes the bandwidth allocation, UAV-user association, and UAV trajectories to enhance the overall service stability. Specifically, the design incorporates a novel utility function based on MPT to achieve a better balance between the sum rate and user rate fluctuations. To solve the joint optimization problem, we propose an iterative algorithm based on alternating optimization (AO) and successive convex approximation (SCA). Simulation results show that our scheme outperforms the other two baselines in terms of sum rate and user rate fluctuations. Furthermore, the resilience requirement in terms of sum rate, user rate fluctuations and user fairness can be achieved by flexibly tuning weight factor in our proposed algorithm.


\end{abstract}


\IEEEpeerreviewmaketitle

\section{Introduction}
The sphere of Unmanned Aerial Vehicle (UAV) communications has witnessed significant advancements in recent times. These advancements are primarily fueled by the UAVs' innate attributes, such as their high maneuverability, mobility, and the favorable line-of-sight (LoS) propagation they facilitate. The transition from single UAV system to multi-UAV system has ushered in a new era of communication, characterized by expansive coverage, heightened communication capacity, and enhanced stability. Recently, there have been extensive studies developed in multi-UAV communication system. In \cite{9877894}, both access control and deployment design have been studied to maximize the throughput. In \cite{9892688}, trajectories of multiple UAVs have been cooperatively designed for throughput maximization while maintaining service fairness. {\color{black}In \cite{wu2018joint}, the UAV  trajectory, power control, and multiuser scheduling \& association are jointly optimized to maximize the minimum throughput over all ground users.}

In many UAV application scenarios, harsh environmental conditions can cause UAV malfunctions, leading to interruption of UAV-user communication. When certain UAVs fail or run out of power, it is vital to utilize the remaining communication resources fully to ensure a good user experience. Nevertheless, the above studies have not considered the risk of communication interruptions caused by unexpected UAV link failures. A limited studies have addressed the issue of UAV link re-establishment \cite{sharma2015self,chen2020sidr,mou2021resilient}. In \cite{sharma2015self}, a self-healing neural model has been proposed, specifically tailored to manage uncertain UAV link failures. Reference \cite{chen2020sidr} has introduced a damage-resilient method to recover the severely damaged UAV network. And \cite{mou2021resilient} has employed a graph convolutional network approach to tackle the self-healing problem. While these studies primarily emphasize link restoration for service continuity, they overlook the system design, which leads to drastic user rate fluctuations and the system sum rate decrease.

To counteract UAV link failures, we propose a resilient multi-UAV network design enriched with the Modern Portfolio Theory (MPT) \cite{markowitz2000mean}, an investment strategy distinguished for its adept risk management. Analogous to risk-averse formulations in MPT, we construct a utility function that can be maximized to achieve the balance between the sum rate and user rate fluctuations through the joint optimization of bandwidth allocation, UAV-user association and UAV trajectories. The formulated problem is challenging to solve due to the intrinsic sequential non-linear characteristic and the coupling mixed integer variables. Here, we leverage the exponential function that captures both mean and variance to transform the original problem into a series of independent optimization problems. Then, an iterative algorithm is proposed based on alternating optimization (AO) and successive convex approximation (SCA) to solve the problem. {\color{black}Simulation results demonstrate that our proposed algorithm generates a higher sum rate compared to the conventional schemes when the weight factor absolute value $\left| \mu  \right|$ is small, and better user fairness when $\left| \mu  \right|$ is large.}


\section{System Model and Problem Formulation}

\subsection{System Model}
\begin{figure}[htbp] 
	\centering
	\includegraphics[width=12cm]{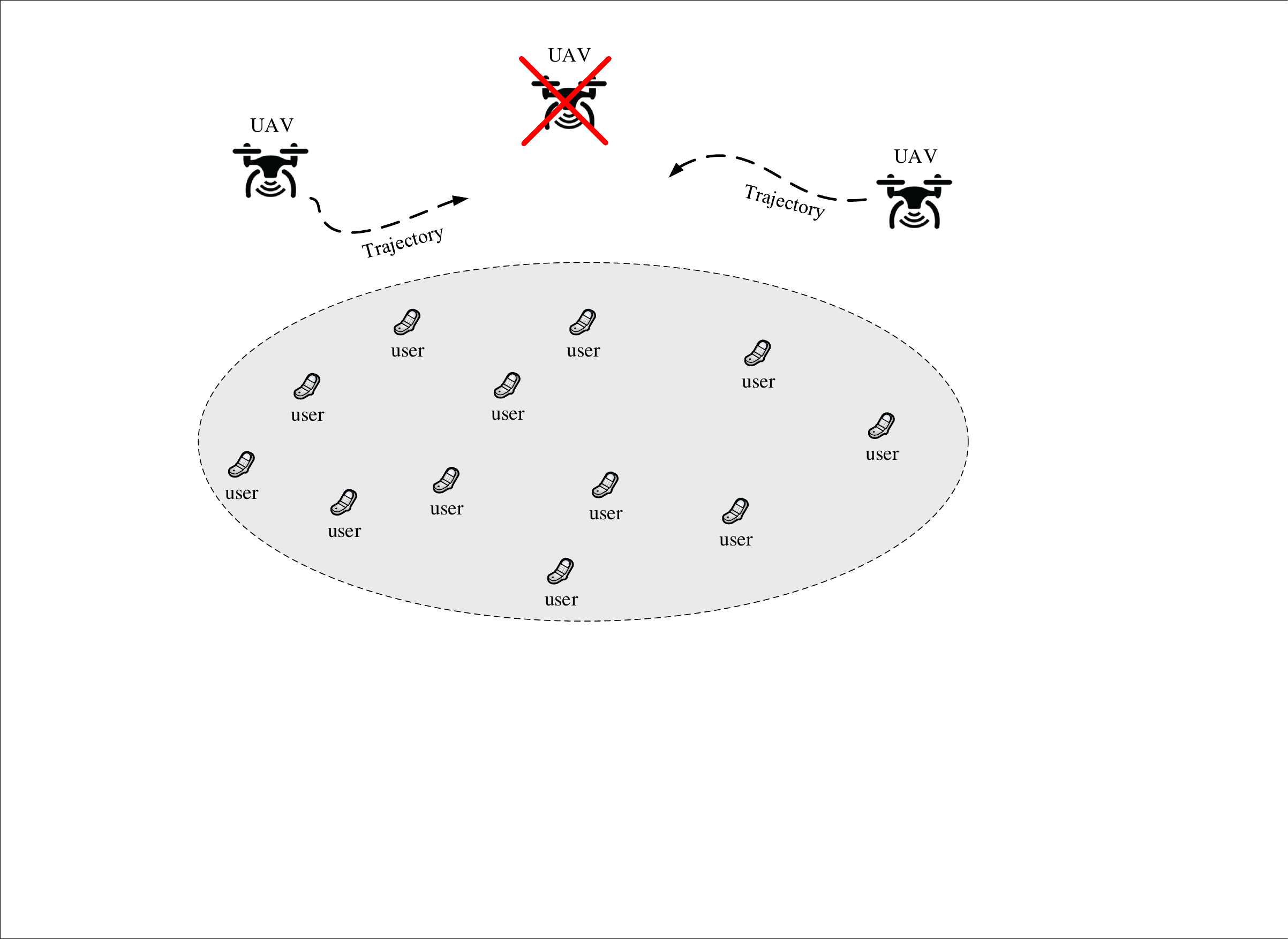}
	\caption{Resilient Multi-UAV Network.}
	\label{figure1}
\end{figure}

As illustrated in Fig. \ref{figure1}, we consider downlink communication between multiple UAVs and a set of quasi-static users, which are denoted by ${\mathcal U} = \left\{ {1, \cdots ,U} \right\}$ and ${\mathcal K} = \left\{ {1, \cdots ,K} \right\}$, respectively. All UAVs are assumed to fly at fixed altitude $H$ to provide communication service for users. The horizontal coordinate for the $u$th UAV at the $n$th time slot is ${{\bf{q}}_u}\left[ n \right]$, while users are located on the ground with horizontal coordinate ${{\bf{c}}_k}$ for the $k$th user. Some of the UAVs fail or run out of power from a certain time slot.

The UAV-user channel is assumed to follow the Rician channel model \cite{yu2021uavs}. Therefore, the channel gain from UAV $u$ to user $k$ at time slot $n$ can be modeled as

\begin{equation}
\begin{aligned}
{h_{u,k}}\left[ n \right] &= \frac{{{\rho}}}{{d_{u,k}^2\left[ n \right]}}{\left| {\sqrt {\frac{{{M}}}{{{M} + 1}}} {{\bar g}_{u,k}}\left[ n \right] + \sqrt {\frac{1}{{{M} + 1}}} {{\tilde g}_{u,k}}\left[ n \right]} \right|^2}\\
&\buildrel \Delta \over = \frac{{{\rho}}}{{{{\left\| {{{\bf{q}}_u}\left[ n \right] - {{\bf{c}}_k}} \right\|}^2} + {H^2}}}{{\bar h}_{u,k}}\left[ n \right],
\end{aligned}
\end{equation}
with ${\rho}$ being the reference channel gain at 1 meter \cite{li2021reconfigurable} and ${d_{u,k}}\left[ n \right]$ being the distance between the UAV $u$ and user $k$. 
\begin{equation}
{\bar h_{u,k}}\left[ n \right] \buildrel \Delta \over = {\left| {\sqrt {\frac{{{M}}}{{{M} + 1}}} {{\bar g}_{u,k}}\left[ n \right] + \sqrt {\frac{1}{{{M} + 1}}} {{\tilde g}_{u,k}}\left[ n \right]} \right|^2},
\end{equation}
where ${M}$ is the Rician factor, and ${\bar g_{u,k}}\left[ n \right]$ and ${\tilde g_{u,k}}\left[ n \right]$ denote the LoS and non-LoS component of the UAV-user link, respectively.

The achievable rate of user $k$ at time slot $n$ is given by

\begin{equation}
\begin{aligned}
{R_k}\left[ n \right] = \sum\limits_{u = 1}^U {{a_{u,k}}\left[ n \right]{r_{u,k}}\left[ n \right]} ,
\end{aligned}
\end{equation}

\noindent where  
\begin{equation}
{r_{u,k}}\left[ n \right] = {b_{u,k}}\left[ n \right]{\log _2}\left( {1 + \frac{{{h_{u,k}}\left[ n \right]{p_u}\left[ n \right]}}{{\sum\limits_{i = 1,i \ne u}^U {{h_{i,k}}\left[ n \right]{p_i}\left[ n \right]}  + {\sigma ^2}}}} \right).
\end{equation}
with ${b_{u,k}}\left[ n \right]$ being the bandwidth allocated to the user $k$ from the UAV $u$ at time slot $n$, ${p_u}\left[ n \right]$ being the transmit power of UAV $u$ at time slot $n$ and ${\sigma ^2}$ being the power of an additive Gaussian noise. ${a_{u,k}}\left[ n \right]$ is the association relation between UAV $u$ and user $k$ at time slot $n$. In detail, ${a_{u,k}}\left[ n \right] = 1 $ if user $k$ is associated to UAV $u$ at time slot $n$ and ${a_{u,k}}\left[ n \right] = 0 $ otherwise.


{\color{black}In MPT \cite{markowitz2000mean}, risk-averse formulations based on Markowitz mean-variance model \cite{markowitz2000mean} are proposed to optimize the allocation of assets in an investment portfolio, which consolidates investment average return and investment risk into a weighted expression. This framework has also been adopted within the wireless communication field \cite{alsenwi2021intelligent}. Similarly, to address the impact of malfunctioning UAVs, we propose a risk-averse formulation that incorporates the variance of user rate and the sum rate.} The user rate refers to the rate of each user, while the sum rate refers to the total rate of all users. The variance of user rate is used to capture the fluctuations of user rate. Specifically, by adopting the risk-averse formulations in MPT, we define a utility function capturing both the average of the sum rate and the variance of user rate as

\begin{equation}
\begin{aligned}
{\mathop{\rm F}\nolimits} \left( {{\bf{A}},{\bf{B,Q}}} \right) = \sum\limits_{k = 1}^K {\left( {\mathbb E{_n}\left[ {{R_k}\left[ n \right]} \right] - \beta {{\mathbb V}_n}\left[ {{R_k}\left[ n \right]} \right]} \right)} ,
\end{aligned}
\end{equation}
where ${\bf{A}} = \left\{ {{a_{u,k}}\left[ n \right],\forall u,k,n} \right\}$ is the UAV-user association matrix, ${\bf{B}} = \left\{ {{b_{u,k}}\left[ n \right],\forall u,k,n} \right\}$ is the bandwidth allocation matrix, and ${\bf{Q}} = \left\{ {{{\bf{q}}_u}\left[ n \right],\forall u,n} \right\}$ is the UAV trajectories matrix, respectively. {\color{black}$\beta$ is the weight for the penalty item that captures the user rate fluctuations brought by UAV link failures. This is similar to the way of investment risk characterization in MPT.}

\subsection{Problem Formulation}
We aim to maximize the utility function by optimizing the UAV-user association, bandwidth allocation and UAV trajectories to reduce the impact of link failures. Accordingly, the optimization problem can be formulated as

\begin{subequations}
	\label{a1}
	\begin{equation}
	\begin{array}{l}
	\mathop {\max }\limits_{{\bf{A,B,Q}}} {\mathop{\rm F}\nolimits} \left( {{\bf{A}},{\bf{B,Q}}} \right) \quad \quad \quad \quad \quad \quad \quad \quad \quad \quad \quad \tag{\ref{a1}}
	\end{array}
	\end{equation}
	\vspace{-2em}
	\begin{align}	
	 {\rm{s}}{\rm{.t}}{\rm{.}}\;&\sum\limits_{u = 1}^U {{a_{u,k}}\left[ n \right]}  = 1,\forall k,n, \label{b1} \\
	 \quad \;&{a_{u,k}}\left[ n \right] \in \left\{ {0,1} \right\},\forall u,k,n, \label{b2} \\
	 \quad \;&\sum\limits_{k = 1}^K {{a_{u,k}}\left[ n \right]{b_{u,k}}\left[ n \right]}  \le {B_u},\forall u,n, \label{b3} \\
	 \quad \;&{\left\| {{\bf{q}}_u\left[ {n + 1} \right] - {\bf{q}}_u\left[ n \right]} \right\|^2} \le D_{\max }^2,\forall u,n, \label{b4} \\
	 \quad \;&{\left\| {{{\bf{q}}_u}\left[ n \right] - {{\bf{q}}_j}\left[ n \right]} \right\|^2} \ge D_{\min }^2,\forall n,u, \ne j \label{b5} \\
	 \quad \;&{{\bf{q}}_{\min }} \le {{\bf{q}}_u}\left[ n \right] \le {{\bf{q}}_{\max }},\forall n,u, \label{b6} \\
	 \quad \;&{{\bf{q}}_u}\left[ 1 \right] = {{\bf{q}}_{0,u}},\forall u, \label{b10}
	\end{align}
\end{subequations}
where constraint (\ref{b1}) ensures that each user can communicate with only one UAV at each time slot, (\ref{b2}) accounts for the fact that $a_{u,k}$ is a binary decision variable and ${B_u}$ in (\ref{b3}) represents the maximum bandwidth of UAV $u$. Besides, (\ref{b4}), (\ref{b6}) and (\ref{b10}) are constraints of UAV mobility and (\ref{b5}) ensures the safety of flying UAVs, where $D_{\max}$, $D_{\min}$ and ${{\bf{q}}_{0,u}}$ are the maximum and minimum horizontal distance at each time slot and the initial position of UAV, respectively. {\color{black}Actually, $D_{\max}$ also limits the maximum speed of UAV. Usually, UAVs with communication tasks might not be allowed to move quickly for the stability of UAV communication network, which means the Doppler effect is insignificant. For larger $D_{\max}$, extra digital signal processing should be adopted to compensate for the potential Doppler frequency shift, which is beyond the scope of this work.} ${{\bf{q}}_{\min }}$ and ${{\bf{q}}_{\max }}$ represent the feasible flying area. Note that the utility function maximization problem is reduced to the normal sum rate maximization problem when $\beta =0$. Furthermore, the overall design is carried out in a sequential style. During the initial period, all UAVs operate normally. However, in a subsequent period, certain UAVs experience malfunction. In contrast to traditional sum rate maximization strategies, our proposed approach carefully allocates resources based on previously achieved rates. This method ensures the effective allocation of resources, mitigating fluctuations of user rates. The specially designed utility function plays a crucial role in maintaining low user rate fluctuations, which in turn contributes to the overall service stability.


\section{Resource Allocation and Trajectory Design}
Clearly, it is hard to directly solve Problem (\ref{a1}) by applying existing optimization techniques. To solve this complicated problem, we first utilize exponential function to reformulate the problem and then adopt AO method by dividing the reformulated problem into three subproblems, which can be solved in an iterative manner.

We consider the exponential function that can capture both the mean and variance of user rate
\begin{equation}
\begin{aligned}
G\left( {{\bf{A}},{\bf{B,Q}}} \right) = \frac{1}{\mu }\log {\mathbb E}_n\left[ {\exp \left( {\mu  {\sum\limits_{k = 1}^K { {{R_k}\left[ n \right]} } } } \right)} \right],
\end{aligned}
\end{equation}
where $\mu = - 2 \beta$ controls the desired risk-sensitivity \cite{mihatsch2002risk}. The Taylor expansion of the exponential utility function is given by \cite{mihatsch2002risk}
\begin{equation}
\begin{aligned}
&G\left( {{\bf{A}},{\bf{B,Q}}} \right) = {{\mathbb E}_n}\left[ {\sum\limits_{k = 1}^K {{R_k}\left[ n \right]} } \right] + \frac{\mu }{2}{{\mathbb V}_n}\left[ {\sum\limits_{k = 1}^K {{R_k}\left[ n \right]} } \right]   + {\cal O}\left( {{\mu ^2}} \right).
\end{aligned}
\end{equation}

Therefore, Problem (\ref{a1}) can be reformulated as
\begin{subequations}
	\label{a2}
	\begin{equation}
	\begin{array}{l}
	\mathop {\max }\limits_{{\bf{A}},{\bf{B,Q}}} \frac{1}{\mu }\log {\mathbb E}{_n}\left[ {\exp \left( {\mu  {\sum\limits_{k = 1}^K {{R_k}\left[ n \right]} } } \right)} \right] \tag{\ref{a2}}
	\end{array}
	\end{equation}
	\vspace{-1.2em}
	\begin{align}	
	{\rm{s}}{\rm{.t}}{\rm{.}}\;& \rm (\ref{b1})-(\ref{b10}). \nonumber \quad \quad \quad \quad \quad \quad \quad \quad 
	\end{align}
\end{subequations}

Compared with the original sequential problem defined in (\ref{a1}), the reformulated exponential utility function can be optimized independently in each period.
 

\subsection{UAV-user Association Problem}
For given ${\bf{B}}$ and ${\bf{Q}}$, the subproblem of optimizing the UAV-user association ${\bf{A}}$ can be expressed as
\begin{subequations}
	\label{a3}
	\begin{equation}
	\begin{array}{l}
	\mathop {\max }\limits_{\bf{A}} \frac{1}{\mu }\log {\mathbb E}{_n}\left[ {\exp \left( {\mu  {\sum\limits_{k = 1}^K {{R_k}\left[ n \right]} } } \right)} \right]
	\tag{\ref{a3}}
	\end{array}
	\end{equation}
	\vspace{-1.2em}
	\begin{align}	
	{\rm{s}}{\rm{.t}}{\rm{.}}\;\;& \rm (\ref{b1}),(\ref{b2}),(\ref{b3}). \nonumber \quad \quad \quad \quad \quad \quad 
	\end{align}
\end{subequations}

{\color{black}The formulated problem can be solved efficiently by different methods including the cross-entropy based approach \cite{huang2018learning} and the variable relaxation approach, etc. We tend to relax the association variable ${a_{u,k}}\left[ n \right]$ to a continuous one ${{\tilde a}_{u,k}}\left[ n \right]$ satisfying $0 \le {{\tilde a}_{u,k}}\left[ n \right] \le 1$.} Therefore, the resulting problem is
\begin{subequations}
	\label{a4}
	\begin{equation}
	\begin{array}{l}
	\quad \mathop {\max }\limits_{{\bf{\tilde A}}} \;\frac{1}{\mu }\log {\mathbb E}{_n}\left[ {\exp \left( {\mu {\sum\limits_{k = 1}^K {{R_k}\left[ n \right]} } } \right)} \right] \quad \quad 
	\tag{\ref{a4}}
	\end{array}
	\end{equation}
	\vspace{-1.2em}
	\begin{align}	
	{\rm{s}}{\rm{.t}}{\rm{.}}\quad &\sum\limits_{u = 1}^U {{{\tilde a}_{u,k}}\left[ n \right]} = 1,\forall k,n\\
	\quad \quad &0 \le {{\tilde a}_{u,k}}\left[ n \right] \le 1,\forall u,k,n\\
	\quad \quad &\sum\limits_{k = 1}^K {{{\tilde a}_{u,k}}\left[ n \right]{b_{u,k}}\left[ n \right]}  \le {B_u},\forall u,n,
	\label{a4a}
	\end{align}
\end{subequations}
where ${\bf{\tilde A}} = \left\{ {{{\tilde a}_{u,k}}\left[ n \right],\forall u,k,n} \right\}$. Based on the following {\bf{Lemma \ref{lemma1}}}, Problem (\ref{a4}) is convex, which can be solved by standard convex optimization solvers, such as the interior point method \cite{boyd2004convex}.

\newtheorem{lemma}{Lemma}
\begin{lemma} \label{lemma1}
	$\frac{1}{\mu }\log {E_n}\left( {\exp \left( {\mu {x_1}} \right) +  \cdots  + \exp \left( {\mu {x_n}} \right)} \right)$ is a concave function with ${\mu <0}$.
\end{lemma}

\newtheorem*{proof}{Proof:}
\begin{proof}
	We define $g\left( {\bf{x}} \right) = \left[ {{\mu {x_1}},{\mu {x_2}}, \cdots ,{\mu {x_n}}} \right]$ and $f\left( {\bf{x}} \right) = \log \left( {{e^{{x_1}}} + {e^{{x_2}}} +  \cdots  + {e^{{x_n}}}} \right)$. It can be proved that $f\left( {\bf{x}} \right)$ is convex and non-decreasing with respect to ${\bf{x}} = \left[ { {{x_1}},{{x_2}}, \cdots ,{{x_n}} } \right]$. Therefore, composite function $f\left( {g\left( x \right)} \right)$ is convex by exploiting compound properties of convex function. Notice that $\mu  < 0$, so we have $\frac{1}{\mu }\log {E_h}\left( {\exp \left( {\mu {x_1}} \right) +  \cdots  + \exp \left( {\mu {x_n}} \right)} \right)$ to be a concave function.	$\hfill\blacksquare$
\end{proof}


\subsection{Bandwidth Allocation Problem}
For given ${\bf{A}}$ and ${\bf{Q}}$, the subproblem to optimize bandwidth allocation ${\bf{B}}$ is given by

\setcounter{equation}{11}
\begin{subequations}
	\label{a5}
	\begin{equation}
	\begin{array}{l}
	\mathop {\max }\limits_{{\bf B}} \;\frac{1}{\mu }\log {\mathbb E}{_n}\left[ {\exp \left( {\mu  {\sum\limits_{k = 1}^K {{R_k}\left[ n \right]} } } \right)} \right]
	\tag{\ref{a5}}
	\end{array}
	\end{equation}
	\vspace{-1.2em}
	\begin{align}	
	{\rm{s}}{\rm{.t}}{\rm{.}}\;\; \rm (\ref{b3}), \quad \quad \quad \quad \quad \quad \quad \quad \quad \quad \nonumber
	\label{a5a}
	\end{align}
\end{subequations}
where the objective function can be proved to be concave similar to {\bf {Lemma} \ref{lemma1}}. Combining with the convex constraint (\ref{b3}), Problem (\ref{a5}) can be solved by the interior point method.


\subsection{Design of UAV trajectories}
For given ${\bf{A}}$ and ${\bf{B}}$, the subproblem to optimize UAV trajectories ${\bf{Q}}$ can be expressed as
\begin{subequations}
	\label{a6}
	\begin{equation}
	\begin{array}{l}
	\mathop {\max }\limits_{{\bf Q}} \;\frac{1}{\mu }\log {\mathbb E}{_n}\left[ {\exp \left( {\mu  {\sum\limits_{k = 1}^K {{R_k}\left[ n \right]} } } \right)} \right] \quad \quad 
	\tag{\ref{a6}}
	\end{array}
	\end{equation}
	\vspace{-1.2em}
	\begin{align}	
	{\rm{s}}{\rm{.t}}{\rm{.}}\;\; \rm (\ref{b4}), (\ref{b5}), (\ref{b6}), (\ref{b10}) \nonumber \quad \quad  \quad \quad  \quad \quad
	\label{a6a}
	\end{align}
\end{subequations}
where the item in ${R_k}\left[ n \right]$ can be rewritten as

\begin{equation}
\begin{aligned}
&{\log _2}\left( {1 + \frac{{{h_{u,k}}\left[ n \right]{p_u}\left[ n \right]}}{{\sum\limits_{i = 1,i \ne u}^U {{h_{i,k}}\left[ n \right]{p_i}\left[ n \right]}  + {\sigma ^2}}}} \right) \\
&\quad \quad = {\log _2}\left( {\sum\limits_{i = 1}^U {{\rho}d_{i,k}^{ - 2}\left[ n \right]{{\bar h}_{i,k}}\left[ n \right]{p_i}\left[ n \right]}  + {\sigma ^2}} \right) \\
&\quad \quad - {\log _2}\left( {\sum\limits_{i = 1,i \ne u}^U {{\rho}d_{i,k}^{ - 2}\left[ n \right]{{\bar h}_{i,k}}\left[ n \right]{p_i}\left[ n \right]}  + {\sigma ^2}} \right).
\end{aligned}
\end{equation}

To solve this subproblem, we first introduce auxiliary variables $\eta _{i,k}^{{\rm{lower}}}\left[ n \right]$ and $\eta _{i,k}^{{\rm{upper}}}\left[ n \right]$ and then SCA method is utilized to solve the transformed problem. {\color{black}The SCA technique \cite{yu2021uavs}\cite{li2021reconfigurable} is an iterative approach that solves non-convex optimization problems by breaking them down into a series of convex subproblems.} Specifically, auxiliary variables $\eta _{i,k}^{{\rm{lower}}}\left[ n \right]$ and $\eta _{i,k}^{{\rm{upper}}}\left[ n \right]$ are introduced as
\setcounter{equation}{14}
\begin{equation}
\eta _{i,k}^{{\rm{lower}}}\left[ n \right] \le d_{i,k}^{ - 2}\left[ n \right] \le \eta _{i,k}^{{\rm{upper}}}\left[ n \right],\forall i,k,n,
\end{equation}
where $\eta _{i,k}^{{\rm{lower}}}\left[ n \right]$ is the lower bound of $d_{i,k}^{ - 2}\left[ n \right]$ and $\eta _{i,k}^{{\rm{upper}}}\left[ n \right]$ is the upper bound of $d_{i,k}^{ - 2}\left[ n \right]$. The item ${\log _2}\left( {1 + \frac{{{h_{u,k}}\left[ n \right]{p_u}\left[ n \right]}}{{\sum\limits_{i = 1,i \ne u}^U {{h_{i,k}}\left[ n \right]{p_i}\left[ n \right]}  + {\sigma ^2}}}} \right)$ can be further transformed as 

	\begin{equation}
\label{d1}
\begin{aligned}
&{\log _2}\left( {1 + \frac{{{h_{u,k}}\left[ n \right]{p_u}\left[ n \right]}}{{\sum\limits_{i = 1,i \ne u}^U {{h_{i,k}}\left[ n \right]{p_i}\left[ n \right]}  + {\sigma ^2}}}} \right)  \\
&\ge {\log _2}\left( {\sum\limits_{i = 1}^U {{\rho}\eta _{i,k}^{{\rm{lower}}}\left[ n \right]{{\bar h}_{i,k}}\left[ n \right]{p_i}\left[ n \right]}  + {\sigma ^2}} \right) - {\log _2}\left( {\sum\limits_{i = 1,i \ne u}^U {{\rho}\eta _{i,k}^{{\rm{upper}},l}\left[ n \right]{{\bar h}_{i,k}}\left[ n \right]{p_i}\left[ n \right]}  + {\sigma ^2}} \right)\\
& \quad  - \frac{{\sum\limits_{i = 1,i \ne u}^U {{\rho}{{\bar h}_{i,k}}\left[ n \right]{p_i}\left[ n \right]\left( {\eta _{i,k}^{{\rm{upper}}}\left[ n \right] - \eta _{i,k}^{{\rm{upper}},l}\left[ n \right]} \right)} }}{{\left( {\sum\limits_{i = 1,i \ne u}^U {{\rho}\eta _{i,k}^{{\rm{upper}},l}\left[ n \right]{{\bar h}_{i,k}}\left[ n \right]{p_i}\left[ n \right]}  + {\sigma ^2}} \right)\ln 2}} \buildrel \Delta \over = {f_{\bm \eta} },
\end{aligned}
\end{equation}
where ${\bm{\eta }} \buildrel \Delta \over =  \left( {\eta _{i,k}^{{\rm{lower}}}\left[ n \right],\eta _{i,k}^{{\rm{upper}}}\left[ n \right]} \right)$.

Accordingly, we have the following new constraints
\setcounter{equation}{16}
\begin{equation}
{\left\| {{{\bf{q}}_i}\left[ n \right] - {{\bf{c}}_k}} \right\|^2} + {H^2} \le \frac{1}{{\eta _{i,k}^{{\rm{lower}}}\left[ n \right]}}, \label{c2}
\end{equation}
\begin{equation}
{\left\| {{{\bf{q}}_i}\left[ n \right] - {{\bf{c}}_k}} \right\|^2} + {H^2} \ge \frac{1}{{\eta _{i,k}^{{\rm{upper}}}\left[ n \right]}}. \label{c3}
\end{equation}

Based on the first-order Taylor expansion, constraints (\ref{c2}) and (\ref{c3}) can be further transformed into
\begin{equation}
\label{b7}
\begin{aligned}
&\frac{1}{{\eta _{i,k}^{{\rm{lower}},l}\left[ n \right]}} - \frac{1}{{{{\left( {\eta _{i,k}^{{\rm{lower}},l}\left[ n \right]} \right)}^2}}}\left( {\eta _{i,k}^{{\rm{lower}}}\left[ n \right] - \eta _{i,k}^{{\rm{lower}},l}\left[ n \right]} \right)  \ge {\left\| {{{\bf{q}}_i}\left[ n \right] - {{\bf{c}}_k}} \right\|^2} + {H^2},
\end{aligned}
\end{equation}
\begin{equation}
\label{b8}
\begin{aligned}
& {\left\| {{\bf{q}}_i^l\left[ n \right] - {{\bf{c}}_k}} \right\|^2} + 2{\left( {{\bf{q}}_i^l\left[ n \right] - {{\bf{c}}_k}} \right)^T}\left( {{{\bf{q}}_i}\left[ n \right] - {\bf{q}}_i^l\left[ n \right]} \right) + {H^2} \ge \frac{1}{{\eta _{i,k}^{{\rm{upper}}}\left[ n \right]}}. 
\end{aligned}
\end{equation}

Similarly, constraint (\ref{b5}) can be replaced by
\begin{equation}
\label{b9}
\begin{aligned}
 &- {\left\| {{\bf{q}}_u^l\left[ n \right] - {\bf{q}}_j^l\left[ n \right]} \right\|^2} + 2{\left( {{\bf{q}}_u^l\left[ n \right] - {\bf{q}}_j^l\left[ n \right]} \right)^T}\left( {{{\bf{q}}_u}\left[ n \right] - {{\bf{q}}_j}\left[ n \right]} \right) \ge D_{\min }^2,\forall n,u \ne j.
 \end{aligned}
\end{equation}

Therefore, problem (\ref{a6}) is approximated as the following problem
\begin{subequations}
	\label{a7}
	\begin{equation}
	\begin{array}{l}
	\mathop {\max }\limits_{{\bf{Q}},{\bm{\eta }}} \;\frac{1}{\mu }\log {\mathbb E}{_n}\left[ {\exp \left( {\mu {\sum\limits_{k = 1}^K { {\sum\limits_{u = 1}^U {{a_{u,k}}\left[ n \right]{b_{u,k}}\left[ n \right]{f_{\bm \eta} }} } } } } \right)} \right] \quad \quad \quad \quad \quad \quad \quad \quad
	\tag{\ref{a7}}
	\end{array}
	\end{equation}
	\vspace{-2em}
	\begin{align}	
	\quad 	{\rm{s}}{\rm{.t}}{\rm{.}}\quad & \rm (\ref{b4}), (\ref{b6}), (\ref{b10}), (\ref{b7}), (\ref{b8}), (\ref{b9}). \nonumber \quad \quad \quad \quad \quad \quad \quad \quad \quad \quad
	\end{align}
\end{subequations}

\noindent which is a convex problem and can be solved by interior point method. 

The overall algorithm is summarized in Algorithm \ref{alg:alg1}. The objective function is non-decreasing after each iteration of the overall algorithm and upper bounded by a finite value. Thus, the proposed algorithm is guaranteed to converge. {\color{black}When $\beta$ is equal to 0, the original optimization problem can be simplified to a sum rate maximization problem, which shares the partial objective function of the utility function maximization problem. The solution process for the sum rate maximization problem is slightly different from that of the utility function maximization problem, as it does not require the use of exponential functions to transform the objective function. Instead, each subproblem can be solved similarly to Sections A, B, and C, since we have proved that the function $\frac{1}{\mu }\log {E_n}\left( {\exp \left( {\mu {x_1}} \right) +  \cdots  + \exp \left( {\mu {x_n}} \right)} \right)$ has convexity-preserving properties.}

\begin{algorithm}[h]
	\caption{\label{alg:alg1}Resource Allocation and Trajectory Design} 
	\begin{algorithmic}[1]
		\State \textbf{Initialization}
		\State Initialize the iteration index $l=1$, the maximum number of iterations $L$, UAV-user association ${\bf{{\tilde A}}}^{\left( l \right)}$, bandwidth allocation ${\bf{B}}^{\left( l \right)}$ and UAV trajectories ${\bf{Q}}^{\left( l \right)}$.
		\State \textbf{repeat}
		\State With given ${\bf{B}}^{\left( l \right)}$ and ${\bf{Q}}^{\left( l \right)}$, obtain ${\bf{{\tilde A}}}^{\left( l+1 \right)}$ by solving (\ref{a4});
		\State With given ${\bf{{\tilde A}}}^{\left( l+1 \right)}$ and ${\bf{Q}}^{\left( l \right)}$, obtain ${\bf{B}}^{\left( l+1 \right)}$ by solving (\ref{a5});
		\State With given ${\bf{{\tilde A}}}^{\left( l+1 \right)}$ and ${\bf{B}}^{\left( l+1 \right)}$, obtain ${\bf{Q}}^{\left( l+1 \right)}$ by solving (\ref{a7});
		\State $l \leftarrow l + 1$;
		\State \textbf{until} convergence
		\State Map ${\bf{\tilde A}}$ into the binary one ${\bf{A}}$ \cite{alsenwi2021intelligent}.
	\end{algorithmic}
\end{algorithm}


\section{Simulation Results}
In this section, simulation results are presented to evaluate the performance of our proposed joint resource allocation and trajectory design for resilient multi-UAV communication systems. In our simulation, there are three UAVs flying at the altitude $H = 60\; {\rm m}$, which are initially located at $\left( {125\;{\rm{m}},375\;{\rm{m}}} \right)$, $\left( {375\;{\rm{m}},375\;{\rm{m}}} \right)$ and $\left( {250\;{\rm{m}},125\;{\rm{m}}} \right)$, respectively. One of UAVs fails or runs out of power in midway and the other UAVs are reconfigured to serve all users. The reference channel gain at 1 meter is ${\rho} =  - 20\;{\rm{dB}}$ and the Rician factor is ${M} = 3\;{\rm{dB}}$ \cite{li2021reconfigurable}. The remaining parameters are set as ${B_u} = 10\;{\rm{kHz}}$, ${D_{\max }} = 25\;{\rm{m}}$, ${D_{\min }} = 4\;{\rm{m}}$ and total time slots $N=20$, respectively. 

In our simulation, several schemes are compared. Our proposed Algorithm is denoted as ``Pro-Alg''. {\color{black}It has a special case where $\beta = 0$, which we call ``SR-Max''. This variant focuses solely on maximizing the sum rate and does not take into account the minimization of user rate fluctuations.} The scheme of equal bandwidth allocation where users are associated with the nearest UAV \cite{xu2019cellular} is denoted as ``Baseline 1". The scheme in \cite{nguyen2019uav} is denoted as ``Baseline 2", where the sum rate is maximized through the optimization of UAV placement and bandwidth allocation.

\begin{figure}[htbp] 
	\centering
	\includegraphics[width=10cm]{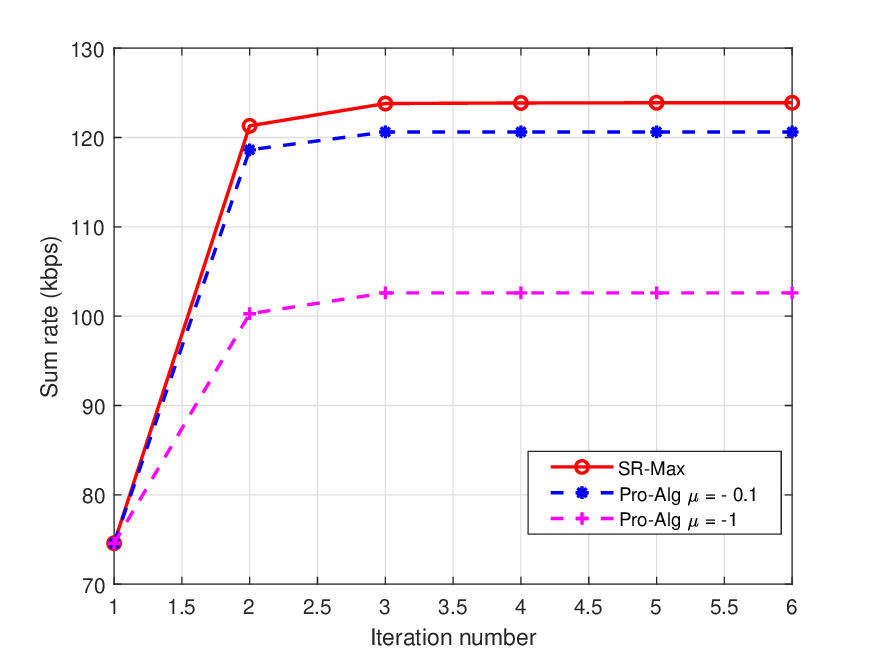}
	\caption{Convergence of the proposed algorithms for $K=9$.}
	\label{figure2}
\end{figure}


The convergence of our proposed Algorithm has been demonstrated under several values of $\mu$ in Fig. \ref{figure2}. Simulations show that sum rates always increase quickly with a few iterations and converge within three iterations.

\begin{figure}[!t] 
	\centering
	\includegraphics[width=10cm]{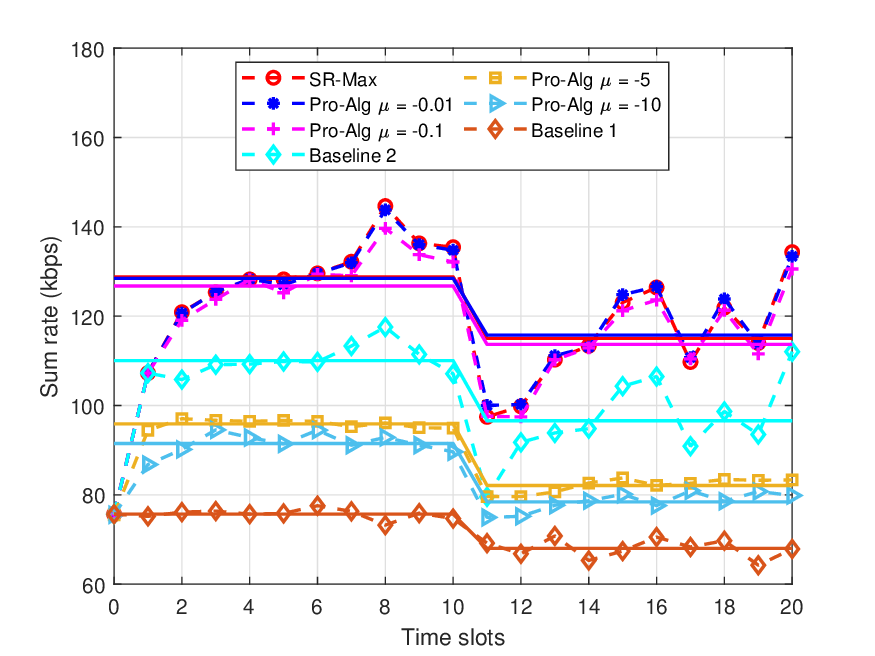}
	\caption{Sum rate for different periods for $K=9$. (Solid line represents average sum rates every 10 time slots)}
	\label{figure3}
\end{figure}


\begin{figure}[!t] 
	\centering
	\includegraphics[width=10cm]{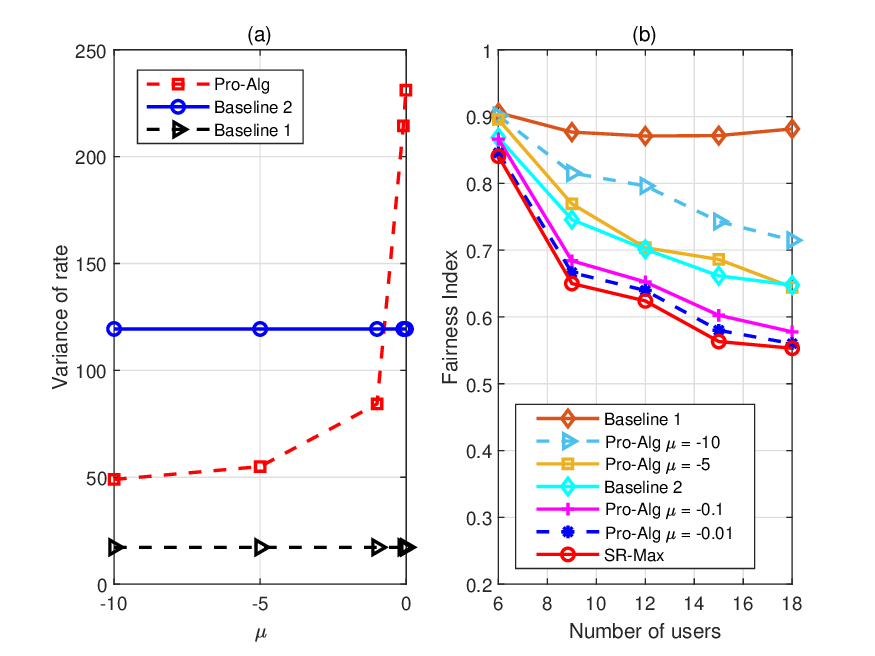}
	\caption{(a) Variance of rate versus $\mu$;  (b) The Jain's fairness among users.}
	\label{figure5}
\end{figure}


Fig. \ref{figure3} illustrates the achieved sum rate comparison for different schemes. Here, we consider two periods. In the first period, three UAVs work normally providing communication services. {\color{black}Upon observation, it is evident that SR-Max (average sum rate = 128.8 kbps) outperforms all other schemes in terms of sum rate. Baseline 1 (average sum rate = 75.68 kbps), which uses a simple association policy, has the worst sum rate. The average sum rate of SR-Max is 70.19\% higher than that of Baseline 1 and 17.09\% higher than that of Baseline 2. Although Baseline 2 also maximizes the sum rate, it achieves a much lower sum rate since it does not take into consideration the trajectories of UAVs or UAV-user association.} 

In the second period, just at the beginning of the 11-th time slot, one of UAVs suddenly breaks down without known by algorithms in advance. The sum rates of all schemes drop inevitably. After that, all schemes try to recover from such accident with timely adjustments. We notice that sum rate fluctuation of our proposed algorithm will decrease with the increase of the absolute value of $\mu$, which shows the effect of the penalty term of sum rate in (8). In order to better illustrate the trade off between the sum rate and its fluctuation, Fig. \ref{figure5}(a) depicts the variance of sum rate of Fig. \ref{figure3} over 20 time slots versus the value of $\mu$. {\color{black}When $\mu = -10$, Pro-Alg achieves a rate variance of 48.97, which is only 41.01\% of that achieved by Baseline 2 (119.4). This observation suggests that a larger absolute value of $\mu$ results in smaller fluctuations in the sum rate. Combined with the sum rate in Fig. \ref{figure3}, enjoying a higher sum rate also entails enduring more pronounced rate fluctuations. This is because a higher priority is given to the variance part, which leads to the allocation of more resources to handle potential unexpected events. Consequently, the value of $\mu$ plays a crucial role in achieving a balance between the sum rate and its fluctuations, enabling the realization of truly resilient UAV networks. In contrast, Baselines 1 and 2 lack flexibility since they do not consider the variance of user data rate caused by malfunctions.}

{\color{black}In Fig. \ref{figure5}(b), the rate fairness of different schemes is compared using the Jain fairness index \cite{lin2022prioritized}. The results show that SR-Max has the worst fairness (0.65), with a fairness index that is 25.8\% lower than that of Baseline 1 (0.876). This can be attributed to the fact that SR-Max allocates more resources to users with better channel states to maximize the sum rate, without taking into account the possibility of potential accidents or failures.} In contrast, the objective function in Pro-Alg $\mu  < 0$ is composed by sum rate and variance of user rate, thus leading to higher fairness. Besides, the larger the absolute value of $\mu$ is, the higher the fairness is, which is affected by the weight of the penalty item. Collectively, another trade-off between sum rate and user fairness can be achieved by our proposed algorithm and the degree of resilience between sum rate and user fairness can also be flexibly controlled through properly tuning the value of $\mu$. {\color{black}Similar to the process depicted in Fig. \ref{figure5}(a), we can scan $\mu$ and invoke our proposed algorithm to find a proper $\mu$ based on the system's specific requirements for sum rate, rate fluctuations, and user rate fairness.}

\section{Conclusion}

This paper proposes a resilient design for multi-UAV communication systems. UAV-user association, bandwidth allocation, and UAV trajectories are jointly optimized to reduce user rate fluctuations and increase the overall sum rate, thereby decreasing the risk induced by UAV link failures. To solve the optimization problem, we propose an iterative algorithm that leverages both AO and SCA techniques. Our simulation results demonstrate that this scheme yields significant performance gains. Furthermore, our scheme can also achieve a better user fairness by tuning the weight factor.


\footnotesize
\bibliographystyle{IEEEtran}
\bibliography{ref}
\end{document}